\documentclass[aps,twocolumn,showpacs, showkeys,nofootinbib,floatfix]{revtex4-1}

\usepackage{amsmath}
\usepackage{amsfonts}
\usepackage{txfonts}
\usepackage{graphicx}
\usepackage{dcolumn}
\usepackage{bbm}
\usepackage{amssymb}
\usepackage{latexsym}
\usepackage{titlesec}
\usepackage[colorlinks=true, linkcolor=red, citecolor=blue]{hyperref}
\usepackage{longtable}

\begin{document}

\title{Reconciling the Tension Between Planck and BICEP2 Through Early Dark Energy}

\author{Lixin Xu$^{1,2}$}
\email{Corresponding author: lxxu@dlut.edu.cn}

\author{Baorong Chang$^{1}$}

\author{Weiqiang Yang$^{1}$}

\affiliation{$^{1}$Institute of Theoretical Physics, School of Physics \&
Optoelectronic Technology, Dalian University of Technology, Dalian,
116024, P. R. China}

\affiliation{$^{2}$State Key Laboratory of Theoretical Physics, Institute of Theoretical Physics, Chinese Academy of Sciences}

\begin{abstract}
We show the possibility that the observational results of the primordial gravitational waves from Planck and BICEP2 for the tensor-to-scalar ratio $r$ can be reconciled when an early dark energy was included. This early dark energy behaves like a radiation component at very early epoch. This is equivalent to induce additional number of effective neutrino species: $\Delta N_{eff}=[\frac{7}{8}(\frac{4}{11})^{4/3}]^{-1}\rho_{de}(a)/\rho_{\gamma}(a)$, where $\rho_{\gamma}(a)$ is the photon energy density and the numerical factors arise from converting to effective neutrino species. And $\rho_{de}(a)$ is the energy density of early dark energy. Combining the Planck temperature data, the WMAP9 polarization data, and the baryon acoustic oscillation data with and without BICEP2 data, we find that in this early dark energy model the tension between the observations from Planck and BICEP2 was relived at $2\sigma$ regions. But it cannot be removed completely due to the small ratio of early dark energy constrained by the other cosmic observations. As a byproduct, the tension between observed values of Hubble parameter from Planck and the direct measurement of the Hubble constant was removed in this early dark energy model. 
\end{abstract}


\maketitle

\section{Introduction}

The Background Imaging of Cosmic Extragalactic Polarization (BICEP2) experiment \cite{ref:BICEP21,ref:BICEP22} has detected the B-modes of polarization in the cosmic microwave background, where the tensor-to-scalar ratio $r=0.20^{+0.07}_{-0.05}$ with $r=0$ disfavored at $7.0\sigma$ of the lensed-$\Lambda$CDM model was found. However, combining with WMAP9 polarization data, ACT and SPT, Planck group reported a much smaller tensor-to-scalar ratio, compared to that from BICEP2 $r<0.11$ at $95\%$ C.L. in the $\Lambda$CDM+$r$ model \cite{ref:Planck2013tensor}. It apparently shows the tension between these two observations. To relieve this tension, some possible extensions to the $\Lambda$CDM+$r$ model have been discussed such as the ruining spectral index \cite{ref:BICEP21} introduced by BICEP2 (see also in Refs. \cite{ref:runMa,ref:runCzery}), the additional relativistic degrees of freedom beyond the three active neutrinos and photons \cite{ref:Zhang,ref:Hu,ref:Li,ref:Anchordoqui,ref:Ko}, the suppression of the adiabatic perturbations on large scales \cite{ref:Contaldi,ref:Miranda,ref:Abazajian,ref:Hazra}, the blue tilted tensor \cite{ref:Gerbino,ref:Ashoorioon,ref:Wu2014}, and the isocurvature mode \cite{ref:Kawasaki1,ref:Kawasaki2} and so on. In this paper, we take a somewhat related approach to investigate the possibility of relaxing the tension by considering early dark energy which mimics radiation at early epoch. Therefore, effective neutrino species \cite{ref:edeLinder}
\begin{equation}
\Delta N_{eff}=\left[\frac{7}{8}\left(\frac{4}{11}\right)^{4/3}\right]^{-1}\rho_{de}(a)/\rho_{\gamma}(a)\approx4.4032\frac{\Omega_{de}(a)}{\Omega_{\gamma}(a)},
\end{equation}  
were induced, where $\rho_{\gamma}(a)$ and $\rho_{de}(a)$ are the photon energy density and dark energy density respectively. If dark energy contributes to the energy budge about $\Omega^{e}_{de}=0.10$ at early time, there is equivalently a half effective neutrino, when one assumes two main components: radiation+early dark energy. This effective neutrino would modify the sound horizon at recombination, then the inferred distance from both CMB and BAO would be changed. Therefore, it is helpful to relax the tension between that with local measurement of the Hubble parameter. We investigate the effects on the tensor-to-scalar ratio and tension relaxation in a quantitative way by performing Markov chain Monte Carlo (MCMC) analysis by using Planck, BAO and BICEP2 data. 

This paper is structured as follows. In Section \ref{sec:ede}, we give a brief review of an early dark energy model. The data sets and constrained results are given in Section \ref{sec:results}. Section \ref{sec:conclusion} is the conclusion.
 
\section{Early Dark Energy Model} \label{sec:ede}

\begin{center}
\begin{figure}[tbh]
\includegraphics[width=8.5cm]{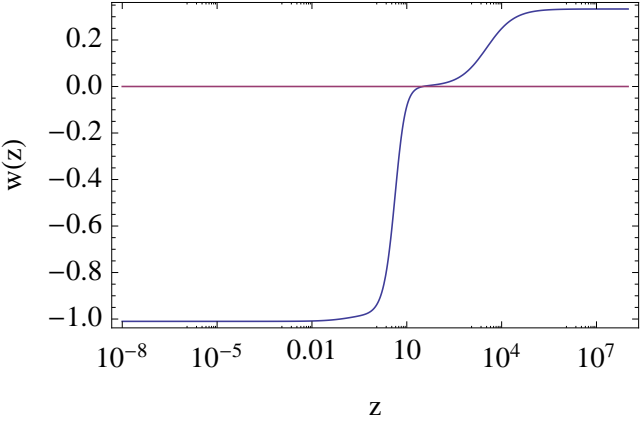}
\caption{The evolution of the EoS of early dark energy with respect to the redshift $z$, where $\Omega_{m}=0.27$, $\Omega^e_{de}=0.04$ and $w_0=-1.01$ were adopted.}\label{fig:eosz}
\end{figure}
\end{center}

The dark energy was introduced to resolve the late time accelerated expansion of our Universe which was discovered in 1998 \cite{ref:Riess98,ref:Perlmuter99}. Usually, a late time negative equation of state (EoS) for dark energy is demanded to provide a repulsive force which pushes our Universe into an accelerated expansion phase. Of course, a modification of gravity theory can also give a late time accelerated expansion at large scales. But, we mainly focus on general relativity, i.e. Einstein gravity theory, in this work. However, this dark sector could have a different character at high redshift, and it can contribute dynamically at early time. Even it acts in a decelerated manner like cold dark matter or radiation \cite{ref:edeLinder,ref:edeDoran}. In this paper, we consider an early dark energy, which behaves like a radiation component at early epoch. We take the one proposed in Ref. \cite{ref:edeDoran} as a simple model. Instead of parameterizing EoS $w(a)$ of dark energy, one can parameterize the dimensionless energy density $\Omega_{d}(a)$ directly from the Friedmann equation
\begin{equation}
H^2(a)=H^2_0\left[\Omega_{m}a^{-3}+\Omega_{\gamma}a^{-4}\right]+H^2(a)\Omega_{de}(a),
\end{equation}        
where $\Omega_{i}=8\pi G\rho_{i,0}/3H^2_0$ is the present dimensionless energy for energy component $\rho_{i}$. Here $\Omega_{m}=\Omega_{b}+\Omega_{c}$, and $\Omega_{\gamma}$ denotes the relativistic energy component. $\Omega_{de}(a)=8\pi G \rho_{de}(a)/3H^2$ is the dark energy which is parameterized as \cite{ref:edeDoran}
\begin{equation}
\Omega_{de}(a)=\frac{\Omega_{de}-\Omega^{e}_{de}(1-a^{-3w_0})}{\Omega_{de}+\Omega_{m}a^{3w_0}+\Omega_{\gamma}a^{3w_0-1}}+\Omega^{e}_{de}(1-a^{-3w_0}),
\end{equation} 
here $\Omega_{de}=1-\Omega_{m}-\Omega_{\gamma}$, $w_0$ is the EoS of dark energy at present, and $\Omega^{e}_{de}$ is the contribution of dark energy at early epoch. From the energy conservation equation for the dark energy, one has the relation between the equation of state and the dimensionless energy density as \cite{ref:Wetterich}
\begin{equation}
\left[3w_{de}(a)-\frac{\Omega_{\gamma}}{\Omega_{m}a+\Omega_{\gamma}}\right]=-\frac{1}{1-\Omega_{de}(a)}\frac{d\ln \Omega_{de}(a)}{d\ln a}.
\end{equation}
An interesting property of this early dark energy is that it behaves like radiation at early epoch and like dark energy at late time. And at the cross-over scale, it behaves like dark matter. For explicitly, we reproduce the evolution of EoS with respect to the redshifts $z$ in Figure \ref{fig:eosz}, see also the Figure 1 in Ref. \cite{ref:edeDoran}. We will show the somehow radiation like component at early epoch is helpful to reconcile the tensor between the observed values of tensor-to-scalar ratio from Planck and that from BICEP2 quantitatively in the next section.  
   
\section{Data Set and Results} \label{sec:results}

To investigate how the tension can be reduced in a quantitative way in the framework of early dark energy model, we perform MCMC analysis by using the following data sets: 

(i) The newly released BICEP2 CMB B-mode data \cite{ref:BICEP21,ref:BICEP22}. It will be denoted by BICEP2.

(ii) The full information of CMB which include the recently released {\it Planck} data sets which include the high-l TT likelihood ({\it CAMSpec}) up to a maximum multipole number of $l_{max}=2500$ from $l=50$, the low-l TT likelihood ({\it lowl}) up to $l=49$ and the low-l TE, EE, BB likelihood up to $l=32$ from WMAP9, the data sets are available on line \cite{ref:Planckdata}. This dat set combination will be denoted by P+W.

(iii) For the BAO data points as 'standard ruler', we use the measured ratio of $D_V/r_s$, where $r_s$ is the co-moving sound horizon scale at the recombination epoch, $D_V$ is the 'volume distance' which is defined as
\begin{equation}
D_V(z)=[(1+z)^2D^2_A(z)cz/H(z)]^{1/3},
\end{equation}
where $D_A$ is the angular diameter distance. The BAO data include $D_V(0.106) = 456\pm 27$ [Mpc] from 6dF Galaxy Redshift Survey \cite{ref:BAO6dF}; $D_V(0.35)/r_s = 8.88\pm 0.17$ from SDSS DR7 data \cite{ref:BAOsdssdr7}; $D_V(0.57)/r_s = 13.62\pm 0.22$ from BOSS DR9 data \cite{ref:sdssdr9}. This data set combination will be denoted by BAO. 

At first, we modified the publicly available {\bf CAMB} \cite{ref:CAMB} code, which calculates the CMB power spectra, to include the early dark energy model. Here the evolution of early dark energy perturbation was considered in the synchronous gauge
\begin{eqnarray}
\dot{\delta}&=&-(1+w)(\theta+\frac{\dot{h}}{2})-3\mathcal{H}(\frac{\delta p}{\delta \rho}-w)\delta,\label{eq:continue}\\
\dot{\theta}&=&-\mathcal{H}(1-3c^2_{s,ad})+\frac{\delta p/\delta \rho}{1+w}k^{2}\delta-k^{2}\sigma\label{eq:euler}
\end{eqnarray}
following the notations of Ma and Bertschinger \cite{ref:MB}, where the definition of the adiabatic sound speed
\begin{equation}
c^2_{s,ad}=\frac{\dot{p}}{\dot{\rho}}=w-\frac{\dot{w}}{3\mathcal{H}(1+w)}
\end{equation}
was used. Here $\mathcal{H}\equiv a'/a=aH$ is the comoving Hubble parameter while $H=\dot{a}/a$ is the physical Hubble parameter. 

\begin{center}
\begin{figure}[tbh]
\includegraphics[width=8.5cm]{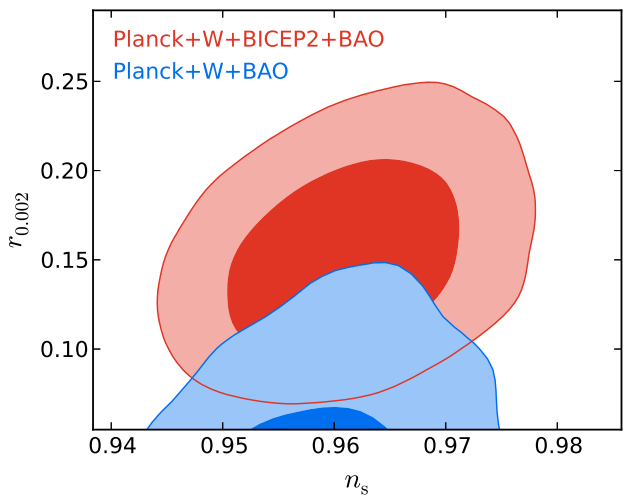}
\caption{The 1D marginalized distribution and 2D contours for interested model parameters with $68\%$ C.L., $95\%$ C.L. by using  {\it Planck} 2013, WMAP9 and BICEP2.}\label{fig:contour}
\end{figure}
\end{center}

We perform a global fitting on the {\it Computing Cluster for Cosmos} by using the publicly available package {\bf CosmoMC} \cite{ref:MCMC} in the following model parameter space
\begin{equation}
P=\{\Omega_b h^2,\Omega_c h^2,  100\theta_{MC}, \tau, n_s, r, {\rm{ln}}(10^{10} A_s),w_0,\Omega^{e}_{de}\},
\end{equation}
their priors are shown in the second column of Table \ref{tab:results}. The running was stopped when the Gelman \& Rubin $R-1$ parameter $R-1 \sim 0.01$ was arrived; that guarantees the accurate confidence limits. The obtained results are shown in Table \ref{tab:results} for the data combinations: Planck+W+BICEP2+BAO and Planck+W+BAO. The obtained contour plots for $r-n_s$ are shown in Figure \ref{fig:contour}. 

\begingroup                                                                                                                     
\squeezetable                                                                                                                   
\begin{center}                                                                                                                  
\begin{table}[tbh]                                                                                                                   
\begin{tabular}{cc@{}|cc@{}|cc}                                                                                                            
\hline\hline                                                                                                                    
Parameters & Priors & \multicolumn{2}{c|}{P+W+BAO} & \multicolumn{2}{c}{P+W+BAO+BICEP2} \\ \hline
$\Omega_b h^2$ & $[0.005,0.1]$ & $0.02193_{-0.00026}^{+0.00027}$ & $0.02187$ & $0.02190_{-0.00026}^{+0.00027}$ & $0.02186$\\
$\Omega_c h^2$ & $[0.001,0.99]$ & $0.1215_{-0.0026}^{+0.0026}$ & $0.1216$ & $0.1208_{-0.0025}^{+0.0025}$ & $0.1204$\\
$100\theta_{MC}$ & $[0.5,10]$ & $1.04071_{-0.00069}^{+0.00069}$ & $1.04079$ & $1.04083_{-0.00066}^{+0.00065}$ & $1.04102$\\
$\tau$ & $[0.01,0.81]$ & $0.088_{-0.014}^{+0.012}$ & $0.087$ & $0.088_{-0.013}^{+0.012}$ & $0.092$\\
$w_0$ & $[-5,0]$ & $-1.18_{-0.11}^{+0.15}$ & $-1.16$ & $-1.16_{-0.11}^{+0.15}$ & $-1.12$\\
$\Omega^e_{de}$ & $[0,0.5]$ & $0.00495_{-0.00495}^{+0.00107}$ & $0.00030$ & $0.00402_{-0.00402}^{+0.00087}$ & $0.00078$\\
$n_s$ & $[0.9,1.1]$ & $0.9585_{-0.0070}^{+0.0071}$ & $0.9570$ & $0.9612_{-0.0067}^{+0.0067}$ & $0.9634$\\
${\rm{ln}}(10^{10} A_s)$ & $[2.7,4]$ & $3.085_{-0.024}^{+0.024}$ & $3.087$ & $3.084_{-0.026}^{+0.023}$ & $3.096$\\
$r_{0.05}$ & $[0,1]$ & $0.043_{-0.043}^{+0.008}$ & $0.00037$ & $0.161_{-0.039}^{+0.032}$ & $0.166$\\
\hline
$\Omega_{de}$ & $...$ & $0.718_{-0.021}^{+0.021}$ & $0.713$ & $0.717_{-0.021}^{+0.021}$ & $0.711$\\
$\Omega_m$ & $...$ & $0.282_{-0.021}^{+0.021}$ & $0.287$ & $0.283_{-0.021}^{+0.021}$ & $0.289$\\
$z_{re}$ & $...$ & $10.95_{-1.10}^{+1.08}$ & $10.96$ & $10.98_{-1.07}^{+1.07}$ & $11.34$\\
$H_0$ & $...$ & $71.67_{-3.43}^{+2.58}$ & $70.90$ & $71.38_{-3.38}^{+2.49}$ & $70.34$\\
$r$ & $...$ & $0.039_{-0.039}^{+0.0064}$ & $0.000325$ & $0.152_{-0.041}^{+0.032}$ & $0.158$\\
${\rm{Age}}/{\rm{Gyr}}$ & $...$ & $13.735_{-0.0547}^{+0.0544}$ & $13.771$ & $13.746_{-0.053}^{+0.053}$ & $13.775$\\
\hline\hline                                                                                                                    
\end{tabular}                                                                                                                                                                                                                                
\caption{The mean values with $1\sigma$ errors and the best fit values of the model parameters and the derived cosmological parameters, where the Planck, WMAP9, BAO with or without BICEP2 data sets were used.}\label{tab:results}                                                                                                 
\end{table}                                                                                                                     
\end{center}                                                                                                                    
\endgroup   

The blue contours show the case of Planck+W+BAO, and the red ones are for the case with the addition of BICEP2. It is clearly to see that in the early dark energy model the values of $r$ is depressed for the data combination Planck+W+BICEP2+BAO due to the fact that the early dark energy mimics the radiation component at early epoch. As a result, it makes the contours overlap in $2\sigma$ region as shown in Figure \ref{fig:contour}. However, in $1\sigma$ regions, there is no any overlaps. Therefore, one can conclude that the including of early dark energy is helpful to relive the tension at $2\sigma$ regions. The main reason is due to the small ratio of early dark energy in the early epoch. If the ratio is large enough at early epoch, the tension would be removed completely as the case of the addition of neutrino. However, the large values are not permitted by the other cosmic observations such as CMB and BAO. As a byproduct, in the early dark energy model, $H_0=71.67_{-3.43}^{+2.58}$ and $H_0=71.38_{-3.38}^{+2.49}$ for Planck+W+BAO and Planck+W+BICEP2+BAO were obtained respectively. Then the tension with the direct measurement of the Hubble constant $H_0=73.8\pm 2.4\text{kms}^{-1}\text{Mpc}^{-1}$ \cite{ref:HST} was removed.

\section{Conclusion} \label{sec:conclusion} 

In this brief paper, we consider the possible relaxation of the tension between the observed values of the tensor-to-scalar ratio $r$ from the Planck and BICEP2 by a special kind of early dark energy model which is characterized by its equation of state. The interesting property is that this early dark energy behaves like radiation early epoch and like dark energy at late time. And a phase transition between them happens around the reionization epoch. Therefore, at early time an effective neutrino species characterized by the ratio of early dark energy component are contributed to the energy budge. And this extra radiation component is helpful to relive the tension between the observations from Planck and BICEP2 at $2\sigma$ region as shown is this paper. However, the tension cannot be removed completely due to the small values of early dark energy contained by other cosmic observations. As a byproduct, the early dark energy model can remove the tension between observed values of Hubble parameter from Planck and the direct measurement of the Hubble constant. In this paper, we did not study the details of the sequences of phase transition of the early dark energy and the physics behind them. It deserves to be studied in the future. We expect this work can shed lights on the discovery of the nature of inflation and dark energy.         

\acknowledgements{After we posted this paper on arXiv, we noticed the relevant papers \cite{ref:Cheng3467,ref:Dinda}. This work is supported in part by NSFC under the Grants No. 11275035 and "the Fundamental Research Funds for the Central Universities" under the Grants No. DUT13LK01.}

\end{document}